\documentclass[twocolumn,preprintnumbers]{revtex4}

\usepackage{graphicx}
\usepackage{epsf}
\usepackage{amsmath,amssymb}
\usepackage{color}
\newcommand{\bvec}{\boldsymbol}

\usepackage{ulem}

\begin{document}
\preprint{KUNS-2785, NITEP 61}
\title{Transition properties of low-lying states in $^{28}$Si probed via 
inelastic proton and alpha scattering}

\author{Yoshiko Kanada-En'yo}
\affiliation{Department of Physics, Kyoto University, Kyoto 606-8502, Japan}
\author{Kazuyuki Ogata} 
\affiliation{Research Center for Nuclear Physics (RCNP), Osaka University,
  Ibaraki 567-0047, Japan}
\affiliation{Department of Physics, Osaka City University, Osaka 558-8585,
  Japan}
\affiliation{
Nambu Yoichiro Institute of Theoretical and Experimental Physics (NITEP),
   Osaka City University, Osaka 558-8585, Japan}

\begin{abstract}
$0^+$, $1^-$, $2^+$, and $3^-$ excitations of $^{28}\textrm{Si}$ 
are investigated via proton and $\alpha$ inelastic scattering off $^{28}\textrm{Si}$.
The structure calculation of $^{28}\textrm{Si}$ is performed with 
the energy variation after total angular momentum and parity projections in the framework of 
antisymmetrized molecular dynamics (AMD).
As a result of the AMD calculation, 
the oblate ground and prolate bands,
$0^+$ and $3^-$ excitations, and the $1^-$ and $3^-$ states of the $K^\pi=0^-$ band are
obtained. Using the 
matter and transition densities of  $^{28}\textrm{Si}$ obtained by AMD, 
microscopic coupled-channel calculations of  proton 
and $\alpha$ scattering off $^{28}\textrm{Si}$ are
performed. The proton-$^{28}\textrm{Si}$ potentials in the reaction calculation
are microscopically
derived by folding the Melbourne $g$-matrix $NN$ interaction with the
AMD densities of $^{28}\textrm{Si}$. 
The $\alpha$-$^{28}\textrm{Si}$ potentials are obtained by folding the 
nucleon-$^{28}\textrm{Si}$ potentials with an  $\alpha$ density. 
The calculation reasonably reproduces the observed elastic and inelastic cross sections of 
proton and $\alpha$ scattering. Transition properties are discussed by combining 
the reaction analysis of 
proton and $\alpha$ scattering and structure features such as transition strengths and form factors.
The isoscalar monopole and dipole transitions are focused.
\end{abstract}

\maketitle

\section{Introduction}

One of interesting phenomena 
concerning nuclear deformations in $sd$-shell nuclei 
is 
shape coexistence of oblate and prolate deformations of $^{28}\textrm{Si}$. 
The ground band of 
the $0^+_1$, $2^+_1$, and $4^+_1$ states was assigned to the oblate band 
from experimental data of the quadrupole moment of the $2^+_1$ state and in-band $E2$ transitions.
The $0^+_2$ state is considered to be a monopole vibration mode on the oblate ground state, 
whereas  the $3^-_1$ state is discussed as an octupole vibration on the ground state.
In addition, the excited $K^\pi=0^+$ band starting from the $0^+_3$ at 6.691 MeV
is considered to be a prolate deformation band. 
On the theoretical side, structure studies of $^{28}\textrm{Si}$ with mean-field \cite{DasGupta:1967nqx} 
 and cluster models  \cite{Bauhoff:1982kd} have suggested 
coexistence of the oblate and prolate shapes, while
modern mean-field calculations failed to describe the prolate $0^+_3$ band 
\cite{Sagawa:2005er,Win:2008vw,Lu:2011wy}. 
In these years, calculations with antisymmetrized molecular dynamics (AMD)
\cite{Kanada-Enyo:2004ere,Taniguchi:2009wp,Chiba:2016zyz}
described the shape coexistence of  $^{28}\textrm{Si}$ and discussed excitations on the oblate 
and prolate deformations. However, assignments of those excited states to experimental
energy levels are remaining issues to be solved. 

In order to clarify deformations and transition properties of these excited states, experiments of 
inelastic electron scattering $(e,e')$ and proton scattering $(p,p')$ off $^{28}\textrm{Si}$
have been performed \cite{Yen:1983sk,Nakada:1972,Sundberg:1967rjo,Horowitz:1969eso,Kato:1985zz,Chen:1990zza}.
The work of Ref.~\cite{Chen:1990zza} 
discussed transition densities from the ground state with reaction analysis of $(p,p')$ data 
at the incident energy $E_p=180$ MeV combined with $(e,e')$ data. 
Detailed studies were performed mainly for strongly populated states, but not done yet for weak transitions.
For example, the $(p,p')$ cross sections  of the $2^+_{1}$ and $3^-_1$ states were
described with a reaction model calculation using  the transition densities reduced from the
charge form factors measured by $(e,e')$ cross sections. 
For consistency check, 
the reduced transition densities were 
found to give consistent values of $B(E2)$ and $B(E3)$  with those 
determined by $\gamma$-decay life times.

For $0^+$ and $1^-$ states, 
there are no  $\gamma$-decay data of the transition strengths.
In principle, the transition strengths can be determined by form factors at low momentum transfer ($q$). 
However, electron scattering data observed for the $0^+_2$ and $1^-$ states 
are not enough to determine precise values of the $E0$ and isoscalar dipole (IS1) transition strengths.
In the experimental study of $(p,p')$ at $E_p=180$ MeV~\cite{Chen:1990zza}, 
a reaction calculation was performed using the transition densities reduced from the $(e,e')$ data 
and succeeded in reproducing the cross sections of the $0^+_2$ state, 
but not the $1^-_1$(8.89 MeV) state.
In the study of $(p,p')$ at $E_p=65$ MeV~\cite{Kato:1985zz}, they tried to describe 
 $1^-_1$(8.89 MeV)  and $3^-_2$(10.18 MeV) cross sections by assuming 
an octupole $K^\pi=0^-$ vibrational band, but the 
calculation failed to reproduce the $1^-_1$(8.89 MeV) data.
For the $0^+_3$(6.691 MeV) of the prolate band, there is almost no data of electron nor proton scattering 
because of weak population in the inelastic scattering. 
 
In these two decades, inelastic $\alpha$ scattering ($\alpha,\alpha')$ has been extensively investigated to obtain information about excited states. Especially, the ($\alpha,\alpha')$ reaction 
has been utilized as a sensitive probe for  
isoscalar monopole (IS0) and IS1 transitions of excited states 
as well as giant resonances in various nuclei
\cite{Harakeh-textbook,Youngblood:1981zz,Clark:1999kan,Youngblood:2001mq,Youngblood:2002mk,John:2003ke,Youngblood:2003jg,Itoh:2003rf,Uchida:2004bs,Lui:2006zk,Youngblood:2007zz,Li:2007bp,Itoh:2013lxs,Youngblood:2015sua,Peach:2016yop,Button:2017osu,Gupta:2018dnh,Adachi:2018pql}. 
It is also a useful tool to search for new cluster states because cluster excitations 
often have strong inelastic transitions  
\cite{Suzuki:1989zza,John:2003ke,Kawabata:2005ta,Wakasa:2006nt,Itoh:2011zz,Yamada:2011ri,Chiba:2015khu}.
Along this line, $(\alpha,\alpha')$ experiments at $E_\alpha=130$ and 386 MeV 
have been performed for various $Z=N$ nuclei in the $sd$-shell region, 
and provided ($\alpha,\alpha')$ cross section data of the $0^+_2$, $0^+_3$, and $1^-$ states of 
$^{28}\textrm{Si}$ \cite{Adachi:2018pql}.  
Now, it is an important issue to investigate transition properties of excited states
with analysis of the ($\alpha,\alpha')$ data combining them with 
$(e,e')$ and $(p,p')$ data as well as $B(E\lambda)$ values determined by $\gamma$ decays.

In our previous studies\cite{Kanada-Enyo:2019prr,Kanada-Enyo:2019qbp}, we have achieved microscopic 
coupled-channel (MCC) calculations of $\alpha$ scattering off $^{12}$C and $^{16}$O,
and succeeded to reproduce the $(\alpha,\alpha')$ cross sections of various excited states
using matter and transition densities of the target nuclei calculated with 
AMD \cite{KanadaEnyo:1995tb,KanadaEn'yo:1998rf,KanadaEn'yo:2012bj}. 
We have performed similar MCC calculations of proton scattering 
for the $2^+_1$ states of various 
$Z\ne N$ nuclei in a light-mass region \cite{Kanada-Enyo:2019uvg}, 
and shown that this approach is applicable for proton and $\alpha$ scattering
off stable and unstable nuclei.

In this study, we apply the MCC approach to $^{28}$Si for calculation of proton and $\alpha$ scattering. 
In our MCC calculations, the Melbourne $g$-matrix $NN$ interaction
are used to construct proton-nucleus and $\alpha$-nucleus potentials in a microscopic folding model (MFM).
An important feature of this effective $NN$ interaction is that  there is no  adjustable parameter
because it was derived based on 
bare nucleon-nucleon interactions.
The original MFM with the 
Melbourne $g$-matrix interaction 
was developed and applied to proton-nucleus elastic scattering successfully in Ref.~\cite{Amos:2000}, 
and its simplified version has been applied systematically to
proton-nucleus \cite{Min10,Toy13,Toyokawa:2015zxa,Minomo:2017hjl} and $\alpha$-nucleus \cite{Egashira:2014zda,Toyokawa:2015zxa} elastic scattering.
Very recently, this framework was applied to MCC calculations of 
 proton and $\alpha$ inelastic processes using the microscopic matter and transition densities obtained by
structure model calculations
~\cite{Minomo:2016hgc,Minomo:2017hjl,Kanada-Enyo:2019prr,Kanada-Enyo:2019qbp,Kanada-Enyo:2019uvg}.

One of the advantages of the present approach is that one can discuss inelastic processes of 
different hadron probes, proton and $\alpha$, in a unified treatment of a microscopic description. 
Another advantage is that there is no adjustable parameter in the reaction part as mentioned above.
Once matter and transition densities are given as structure inputs, one can obtain 
the $(p,p')$ and $(\alpha,\alpha')$ cross sections at given energies without ambiguity. 
Owing to this straightforward procedure from structure inputs to output cross sections, 
one can judge validity of a given structure input via proton and $\alpha$ cross sections 
even if electric data are not  accurate enough to check the input.

In the present paper, we investigate properties of the 
$0^+$, $1^-$, $2^+$, and $3^-$ excitations of $^{28}\textrm{Si}$ 
via inelastic proton and $\alpha$ scattering with the MCC calculation.
A main focus is low-energy IS0 and IS1 excitations from the ground state. 
As for a microscopic description of structure of $^{28}$Si, 
we use an AMD model. 
Since the main concern in this paper is inelastic scattering processes, 
we focus only on the oblate ground band, the lowest prolate bands, and $1^-$ and $3^-$ 
excitations on the oblate state.
In this paper, we start a version of AMD adopted in Ref.~\cite{Kanada-Enyo:2004ere}, 
that is, variation before angular momentum projection with fixed nucleon spins. This version 
was used to describe the oblate and prolate shape coexistence in $N=14$ isotopes including 
$^{28}$Si.
We improve the previous calculation to variation after total angular momentum and 
parity projections (VAP) for calculation of the ground and excited states of $^{28}$Si. 
With the obtained wave functions, we investigate structure properties
such as transition strengths and densities as well as form factors. 
For the use of target densities in the MCC calculation, 
theoretical transition densities obtained by the AMD calculation are renormalized to 
fit experimental data of electric transition strengths and/or charge form factors 
so as to reduce possible ambiguity from the structure model 
as much as possible.

The paper is organized as follows. 
The next section describes frameworks of structure and reaction calculations:
the AMD framework for structure of the target nucleus $^{28}$Si
and the MCC approach for proton-$^{28}$Si and $\alpha$-$^{28}$Si scattering. 
The AMD result for structure properties is shown in Sect.~\ref{sec:results1}, 
and proton and $\alpha$ scattering cross sections obtained by the MCC calculation
are discussed in Sect.~\ref{sec:results2}. 
Combining electric 
properties and hadron inelastic scattering, transition properties of excited states are discussed 
in  Sect.~\ref{sec:discussions}. 
Finally a summary is given in Sect.~\ref{sec:summary}. 

\section{Method} \label{sec:method} 
In this section, the methods of structure and reaction calculations are explained.
For the structure part, a VAP version of AMD is applied to $^{28}$Si.
The reaction calculations of proton and $\alpha$ scattering off $^{28}$Si are performed
with the MCC approach using the AMD densities of $^{28}$Si as 
done in Refs.~\cite{Kanada-Enyo:2019prr,Kanada-Enyo:2019qbp,Kanada-Enyo:2019uvg}.
For details, the reader is referred to the previous works and references therein.

\subsection{AMD calculation for structure of $^{28}$Si}
An AMD wave function of an $A$-nucleon system is given by a Slater determinant of 
single-nucleon Gaussian wave functions as
\begin{eqnarray}
 \Phi_{\rm AMD}({\bvec{Z}}) &=& \frac{1}{\sqrt{A!}} {\cal{A}} \{
  \varphi_1,\varphi_2,...,\varphi_A \},\label{eq:slater}\\
 \varphi_i&=& \phi_{{\bvec{X}}_i}\chi_i\tau_i,\\
 \phi_{{\bvec{X}}_i}({\bvec{r}}_j) & = &  \left(\frac{2\nu}{\pi}\right)^{3/4}
\exp\bigl[-\nu({\bvec{r}}_j-\bvec{X}_i)^2\bigr].
\label{eq:spatial}
\end{eqnarray}
Here ${\cal{A}}$ is the antisymmetrizer, and  $\varphi_i$ is
the $i$th single-particle wave function written by a product of
spatial ($\phi_{{\bvec{X}}_i}$), nucleon-spin ($\chi_i$), and isospin ($\tau_i$) 
wave functions.
In the present calculation of $^{28}$Si, we choose 
 proton up ($p\uparrow$), 
proton down ($p\downarrow$), neutron up ($n\uparrow$), 
neutron down ($n\downarrow$) for the nucleon-spin and isospin wave functions. 
Parameters $\bvec{X}_i$, which describe centroid positions of single-nucleon Gaussian wave packets, 
are treated as variational parameters independently for all nucleons.
This wave function is the same as used in the previous AMD study 
of $^{28}$Si in Ref.~\cite{Kanada-Enyo:2004ere}. 
Using this model wave function, we perform energy variation after 
total-angular-momentum and parity projections (VAP). Namely, parameters $\bvec{X}_i$
for each $J^\pi$ state are determined by energy optimization of 
the $J^\pi$-projected AMD wave function. 

All the parameters for the Gaussian width and effective interactions 
are same as those of Ref.~\cite{Kanada-Enyo:2004ere}
as follows.
The width parameter  $\nu=0.15$ fm$^{-2}$ is used.
For effective nuclear interactions used in the structure calculation, 
the MV1 (case 3) central force \cite{TOHSAKI} supplemented by 
a spin-orbit term of the G3RS force \cite{LS1,LS2} is used.
The Bartlett, Heisenberg, and Majorana parameters
of the MV1 force are $b=h=0$ and $m=0.62$, and the spin-orbit strengths are
$u_{I}=-u_{II}=2800$ MeV.
The Coulomb force is also included.

\subsection{MCC calculation of proton and $\alpha$ scattering off  $^{28}$Si}

Elastic and inelastic cross sections of  proton and $\alpha$ scattering off $^{28}$Si are calculated with
the MCC approach as done in 
Refs.~\cite{Kanada-Enyo:2019prr,Kanada-Enyo:2019qbp,Kanada-Enyo:2019uvg}.
The diagonal and coupling potentials for the nucleon-nucleus system are microscopically calculated
by folding the Melbourne $g$-matrix $NN$ interaction \cite{Amos:2000} with densities of the target nucleus. 
The matter and transition densities of $^{28}$Si obtained by AMD are used as structure inputs for the target nucleus, 

The Melbourne $g$ matrix is an effective interaction containing the density and energy dependences,
which are 
derived by solving a Bethe-Goldstone equation in a uniform nuclear matter with a bare $NN$ interaction of 
the Bonn B potential~\cite{Mac87}.
This interaction was constructed in Ref.~\cite{Amos:2000} and examined 
for systematic investigations of proton elastic and inelastic scattering off various nuclei
at energies from 40~MeV to 300~MeV in Refs.~\cite{Min10,Toy13,Toyokawa:2015zxa,Minomo:2017hjl}.
In the present MCC calculation of proton scattering, \sout{the simplified single-folding model with 
local density approximations is adopted.
the spin-orbit term of the potential is} not taken into account to avoid complexity.

The $\alpha$-nucleus potentials are obtained in an extended nucleon-nucleus
folding (NAF) model \cite{Egashira:2014zda} by folding the nucleon-nucleus potentials 
with an $\alpha$ density. 
For the $\alpha$ density, we adopt the one-range Gaussian distribution given in Ref.~\cite{Satchler:1979ni}.
In the NAF model, energy and density dependences of the 
$g$-matrix $NN$ interaction are taken into account only in the folding process of the target density. 
The validity of the NAF model for $\alpha$ elastic scattering
is discussed in Ref.~\cite{Egashira:2014zda}, and it was successfully applied to 
$\alpha$ inelastic processes in Refs.~\cite{Kanada-Enyo:2019prr,Kanada-Enyo:2019qbp}.

\section{Results of structure calculation of  $^{28}$Si} \label{sec:results1}

The $0^+_1$ state with an oblate shape is obtained by energy variation with the $J^\pi=0^+$ projection. 
$1^-$ and $3^-$ excitations on the oblate ground state
are obtained  by energy variation with the $1^-$, and $3^-$  projections, respectively. 
We label the $1^-$ state as $1^-_\textrm{IS1}$ because of its significant IS1 transition strength.
The $3^-$ state corresponds well to the experimental $3^-_1$(6.879 MeV) state as discussed later, 
and is labeled as $3^-_1$. 
A vibration $0^+$ excitation on the ground state, which we label as $0^+_\textrm{vib}$, 
is obtained by the $0^+$-projected energy variation for orthogonal component to the $0^+_1$ state.
A prolate $0^+$ state is obtained 
as a local minimum with the $0^+$-projected energy variation. 
This state is assigned to the band-head state of the prolate band, and labeled as
$0^+_\textrm{pro}$. 

\begin{figure}[!h]
\includegraphics[width=8.5 cm]{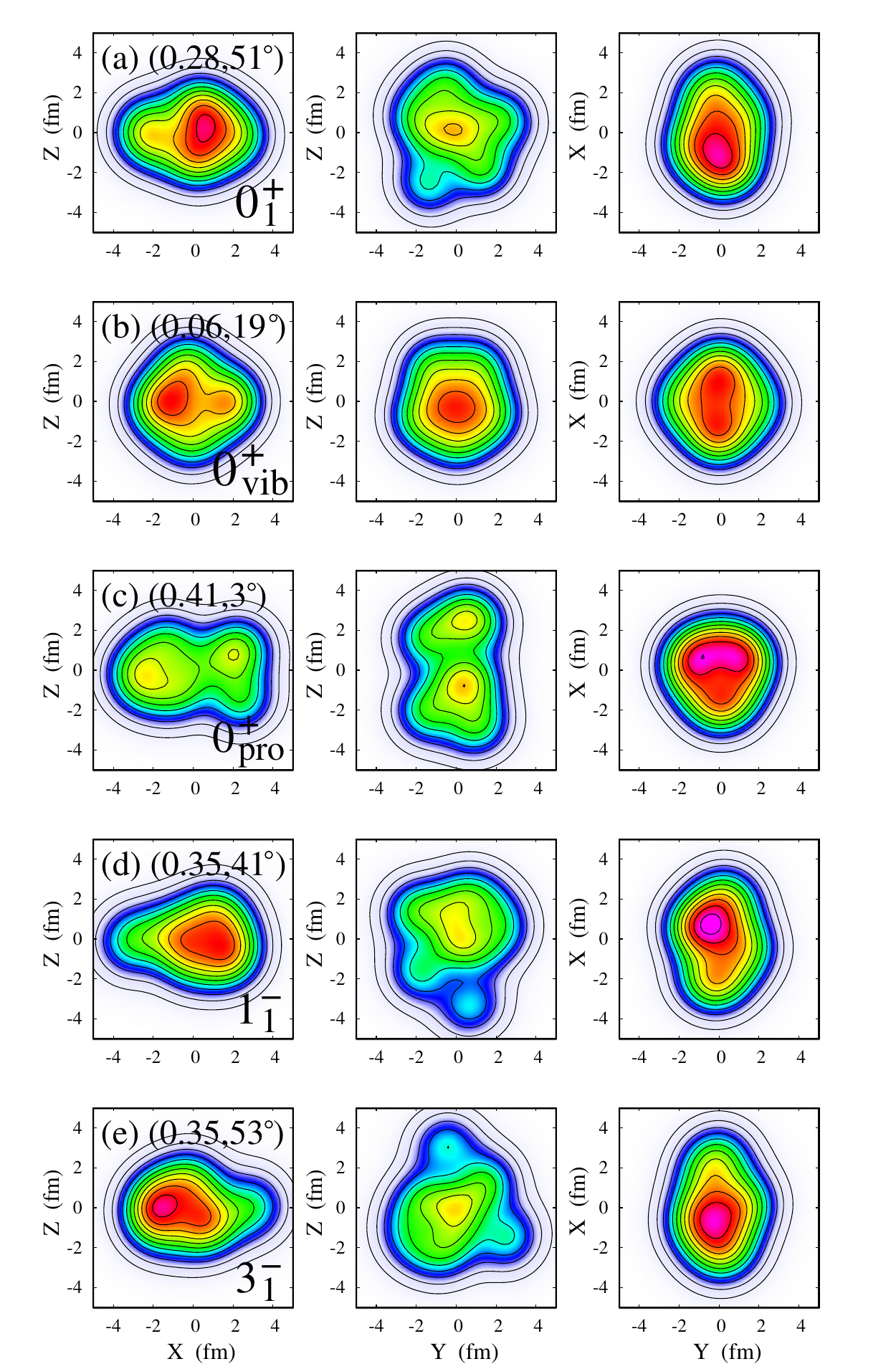}
  \caption{Density distribution of intrinsic wave functions 
for the $0^+_1$, $0^+_\textrm{vib}$, $0^+_\textrm{pro}$, $1^-_\textrm{IS1}$, and $3^-_1$ states 
of $^{28}$Si obtained by AMD.
The density projected onto $X$-$Z$, 
$Y$-$Z$, and $Y$-$X$ planes are shown in left, middle, and right panels, respectively. Here,
intrinsic axises are chosen as 
$\langle ZZ\rangle\ge \langle YY\rangle\ge \langle XX\rangle$ and 
 $\langle XY\rangle=\langle YZ\rangle=\langle ZX\rangle=0$.
The deformation parameters $(\beta,\gamma)$ calculated from the values of 
$\langle ZZ\rangle$,  $\langle YY\rangle$, and $\langle XX\rangle$ are shown in each panel.
  \label{fig:dense-cont}}
\end{figure}

Intrinsic density distributions of the obtained AMD wave functions are shown 
in Fig.~\ref{fig:dense-cont}. The ground state has an approximately
oblate shape with the deformation parameter $\beta=0.28$ (Fig.~\ref{fig:dense-cont} (a)).  
The $0^+_\textrm{vib}$ state is expressed by linear combination of the oblate and 
spherical wave functions shown in Figs.~\ref{fig:dense-cont} (a) and (b), respectively, 
and regarded as a vibration $0^+$ excitation built
on the oblate ground state.
In addition to the oblate state, the prolate deformation with $\beta=0.41$ 
is obtained for the $0^+_\textrm{pro}$ state as shown in  Fig.~\ref{fig:dense-cont} (c).
It constructs a prolate rotational band. 
In the intrinsic density of the $1^-_\textrm{IS1}$ state in Fig.~\ref{fig:dense-cont} (d), 
one can see formation of an $\alpha$ cluster 
at the nuclear surface.  This state is interpreted as 
an $K^\pi=0^-$ excitation mode generated by $\alpha$-cluster motion 
on the oblate state. The  $3^-_1$ state shows a triangle deformation on the oblate state.

In order to calculate energy spectra and wave functions of $J^\pi=0^+,1^-, 2^+,3^-$ and $4^+$ states, we adopt
these five AMD wave functions obtained for $0^+_1$, $1^-_\textrm{IS1}$, $3^-_1$, $0^+_\textrm{vib}$, and 
$0^+_\textrm{pro}$ as basis wave functions to be superposed. 
Namely, we superpose $J^\pi=0^+,1^-, 2^+,3^-$ and $4^+$ eigen states projected from the five basis 
wave functions in order to express $J^\pi_k$ states. 
Coefficients of the linear combination are determined by 
diagonalization of the Hamiltonian and norm matrices. 
As a result of the superposition,  final wave functions for
the $0^+_1$, $1^-_\textrm{IS1}$, $3^-_1$, $0^+_\textrm{vib}$, and 
$0^+_\textrm{pro}$ states and their rotational band members are obtained. 
The binding energy of $^{28}$Si is calculated to be 213.3 MeV,  
which somewhat underestimates the experimental value 236.53 MeV.

\begin{figure}[!h]
\includegraphics[width=8.5 cm]{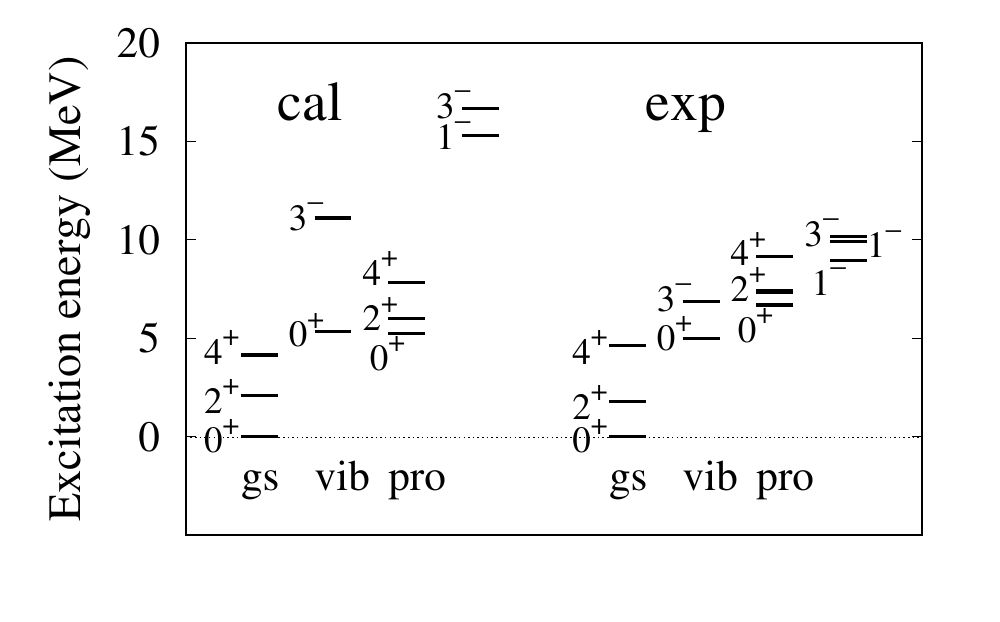}
  \caption{Energy spectra of $^{28}$Si. Left:~Calculated energy spectra of the ground and prolate bands, 
$0^+_\textrm{vib}$ and $3^-_1$ excitations on the ground band, and the $K^\pi=0^-$ band of the 
$1^-_\textrm{IS1}$ and $3^-_2$ states. 
Right:~Experimental spectra corresponding to the theoretical states. 
  \label{fig:spe}}
\end{figure}

\begin{figure}[!h]
\includegraphics[width=6 cm]{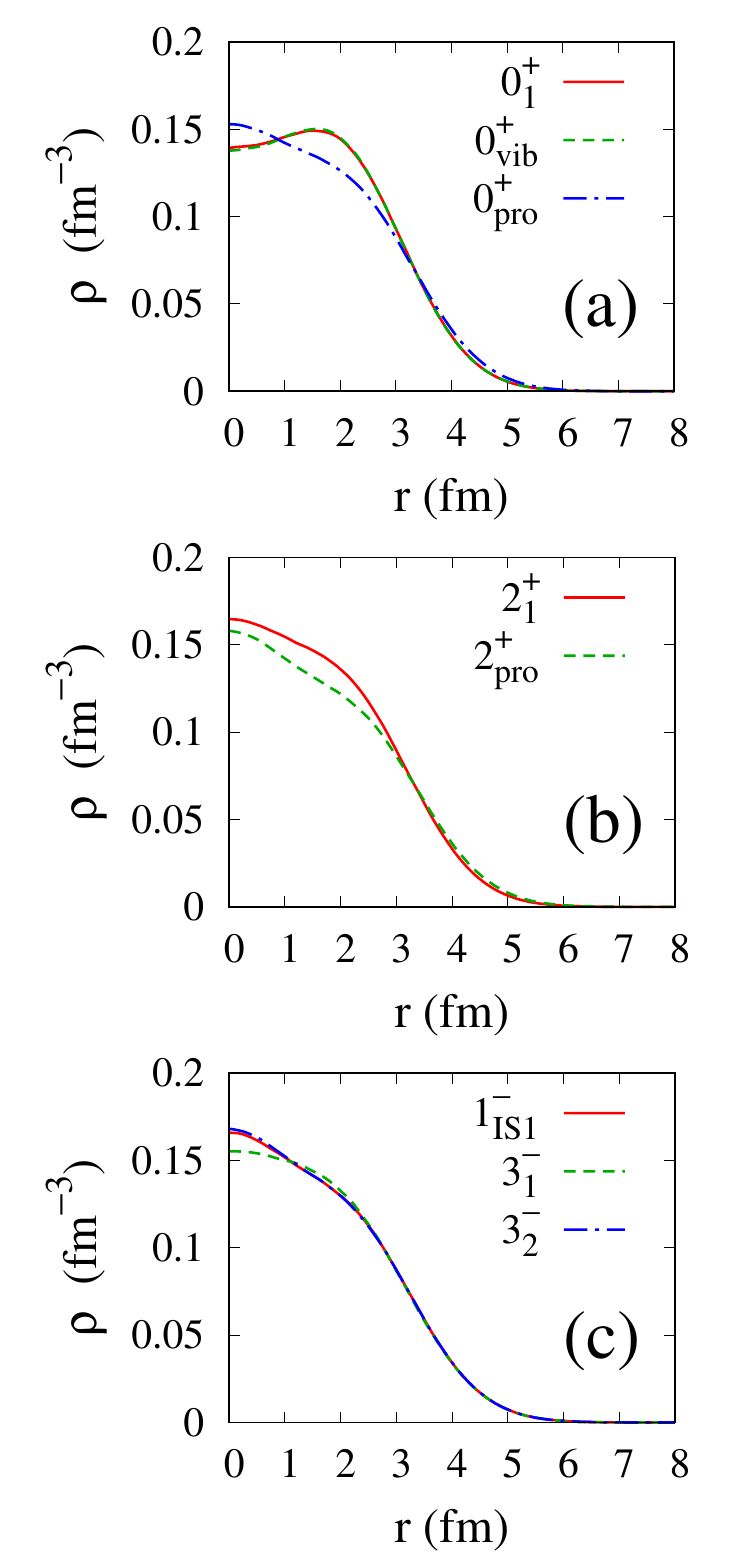}
  \caption{Matter densities of the $0^+$, $2^+$, $1^-$, and $3^-$ states of $^{28}$Si
calculated with AMD.
  \label{fig:density}}
\end{figure}

Based on analyses of intrinsic structure and transition strengths, we classify 
the obtained energy levels into the ground band of the $0^+_1$, $2^+_1$, and $4^+_1$ states, 
the vibration states of the $0^+_\textrm{vib}$ and $3^-_1$ states, 
the prolate band of the $0^+_\textrm{pro}$, $2^+_\textrm{pro}$, and $4^+_\textrm{pro}$ states, 
and the $K^\pi=0^-$ band of the  $1^-_\textrm{IS1}$ and $3^-_2$ states.
The calculated energy spectra are shown in Fig.~\ref{fig:spe} compared 
with the experimental data.
Values of the calculated excitation energies and root-mean-square radii 
as well as experimental energies are listed
in Table \ref{tab:radii}.
The radial distribution of matter density is shown in Fig.~\ref{fig:density}.
Excited states in the prolate and $K^\pi=0^-$ bands have larger matter radii than the
states in the ground band. However, the state dependence of matter 
densities is not so large and may 
give only minor contribution to inelastic scattering of these states.
 
The result of the 
$E0$, $E2$, $E3$, and IS1 transition strengths are listed in Table
\ref{tab:BEl}.
As for the ground band,  the calculation gives strong in-band transitions 
of  $2^+_1\to 0^+_1$ and  $4^+_1\to 2^+_1$ consistently with the experimental $B(E2)$ values. 
For the prolate band, the experimental levels of the $0^+_3(6.691)$, $2^+_2$(7.32 MeV), and 
$4^+_3$(9.16 MeV) are assigned to the rotational band members because 
the $E2$ transition from $4^+_3$(9.16 MeV) is the strongest to the $2^+_2$(7.32 MeV) state. 
However, the observed $E2$ transition from $4^+_3$ is 
fragmented also to the $2^+_3$(7.42 MeV) state, which suggests 
significant state mixing of the prolate $2^+$ state.
In the present calculation, we obtain 
the prolate band members, $0^+_\textrm{pro}$, $2^+_\textrm{pro}$, and $4^+_\textrm{pro}$, 
with almost no fragmentation of the $2^+_\textrm{pro}$ state. 
The calculated $B(E2;4^+_\textrm{pro} \to 2^+_\textrm{pro})$ is consistent 
with a sum of the experimental strengths for the two states, $2^+_2$(7.32 MeV) and $2^+_3$(7.42 MeV). 
As for the $0^+_\textrm{vib}$ state, the calculation gives a slightly higher energy 
than the $0^+_\textrm{pro}$ state. The energy 
ordering of the  $0^+_\textrm{vib}$ and $0^+_\textrm{pro}$ states 
 is not consistent with the 
experimental data of the $0^+_2$ at 4.98 MeV and the $0^+_3$ at 6.69 MeV, which are 
assigned to the vibration and prolate states, respectively. 

The calculated $E0$ transition strength of $0^+_1\to 0^+_\textrm{vib}$ 
is remarkably large and agrees with the experimental $B(E0)$ value for the $0^+_2$(4.98 MeV)
reduced from the $(e,e')$ experiment. On the other hand, for the $0^+_\textrm{pro}$ state, 
we obtain relatively weak $E0$ transition 
because of the shape difference from the ground state. 
It should be commented that the value of 
$B(E0;0^+_1\to 0^+_\textrm{pro})$ is sensitive to the relative energy between the 
$0^+_\textrm{vib}$ and $0^+_\textrm{pro}$ states. 
In the present case, the 
$0^+_\textrm{vib}$ and $0^+_\textrm{pro}$ states almost degenerate with each other. 
This accidental degeneracy somewhat enhances the 
$B(E0;0^+_1\to 0^+_\textrm{pro})$ via the state mixing. 
It means that the predicted value of $B(E0;0^+_1\to 0^+_\textrm{pro})$ may contain 
model ambiguity and should be checked by 
experimental observables of inelastic scattering as discussed later.  

As for the negative-parity states, $1^-_\textrm{IS1}$, $3^-_1$, and $3^-_2$, 
the calculation tends to overestimate the experimental excitation energies, but
it gives reasonable result for inelastic transitions compared with 
$\gamma$-decay and $(e,e')$
data of the $1^-_1$(8.95 MeV),  $3^-_1$(6.88 MeV), 
and  $3^-_2$(10.18 MeV) states.
The calculation obtains the strong $E3$ transition to the $3^-_1$ state
with the triangle shape on the oblate deformation, which is
consistent with the observed $B(E3)$ value of the $3^-_1$(6.88 MeV). 

The $1^-_\textrm{IS1}$ state is characterized by the significant 
IS1 transition, which is induced by 
the $K^\pi=0^-$ excitation mode between mass asymmetric clusters, that is, 
the $\alpha$-cluster motion against the $^{24}$Mg core. It is consistent with the
theoretical work of Ref.~\cite{Chiba:2016zyz} which discussed the 
remarkable IS1 transition of the  $1^-$ state in the $K^\pi=0^-$ band.
The calculated IS1 transition strength of $1^-_\textrm{vib}\to 0^+_1$ is in 
reasonable agreement with the experimental $B(\textrm{IS1})$ value of the  $1^-_1$(8.95 MeV)
reduced from the $(e,e')$ experiment. Therefore, we tentatively assign the 
$1^-_\textrm{IS1}$ state to the  $1^-_1$(8.95 MeV) state. However, 
it should be noted that  
the $1^-_2$(9.93)  state in the experimental spectra 
can be another candidate for the $1^-_\textrm{IS1}$ state because the  IS1 transitions
observed by $(e,e')$ \cite{Yen:1983sk,Chen:1990zza}
 for the $1^-_1$(8.95 MeV) and $1^-_1$(9.93 MeV) states are
almost the same order  as shown later.
The $\alpha$-cluster excitation 
constructs the $K^\pi=0^-$ band consisting of the $1^-_\textrm{IS1}$ and $3^-_2$ states.
The calculated $E3$ transition of $3^-_2\to 0^+_1$ is consistent with
the $(e,e')$ data for the $3^-_2$(10.18 MeV) state.

For the use of the MCC calculation, 
transition densities ($\rho^\textrm{tr}(r)$) are 
calculated with the obtained AMD wave functions. 
In order to reduce model ambiguity from the structure calculation, 
the obtained transition densities are renormalized 
by adjusting the calculated $E\lambda$ transition strength $B_\textrm{th}(E\lambda)$ 
to the observed strength $B_\textrm{exp}(E\lambda)$ 
as  $\rho^\textrm{tr}(r)\to f^\textrm{tr}\rho^\textrm{tr}(r)$ 
with the factor $f^\textrm{tr}=\sqrt{B_\textrm{exp}(E\lambda)/B_\textrm{th}(E\lambda)}$.
The renormalization factors are determined
for the $2^+_1\to 0^+_1$, $2^+_1\to 0^+_1$, $0^+_2\to 2^+_1$, 
$0^+_3\to 2^+_1$, and $3^-_1\to 0^+_1$ transitions with
the $B_\textrm{exp}(E\lambda)$ values of $\gamma$-decay life times, and  $0^+_2\to 0^+_1$ and  $1^-_1\to 0^+_1$
with the $B_\textrm{exp}(E\lambda)$ values reduced by the $(e,e')$ experiment. 
The adopted values of $f^\textrm{tr}$ are listed in Table~\ref{tab:BEl}.
For the $3^-_2\to 0^+_1$ transition, $f^\textrm{tr}=1.40$ 
is chosen so as to fit the charge form factors.
For other transitions, the original transition densities are used as is without renormalization. 
 
\begin{figure}[!h]
\includegraphics[width=8.5 cm]{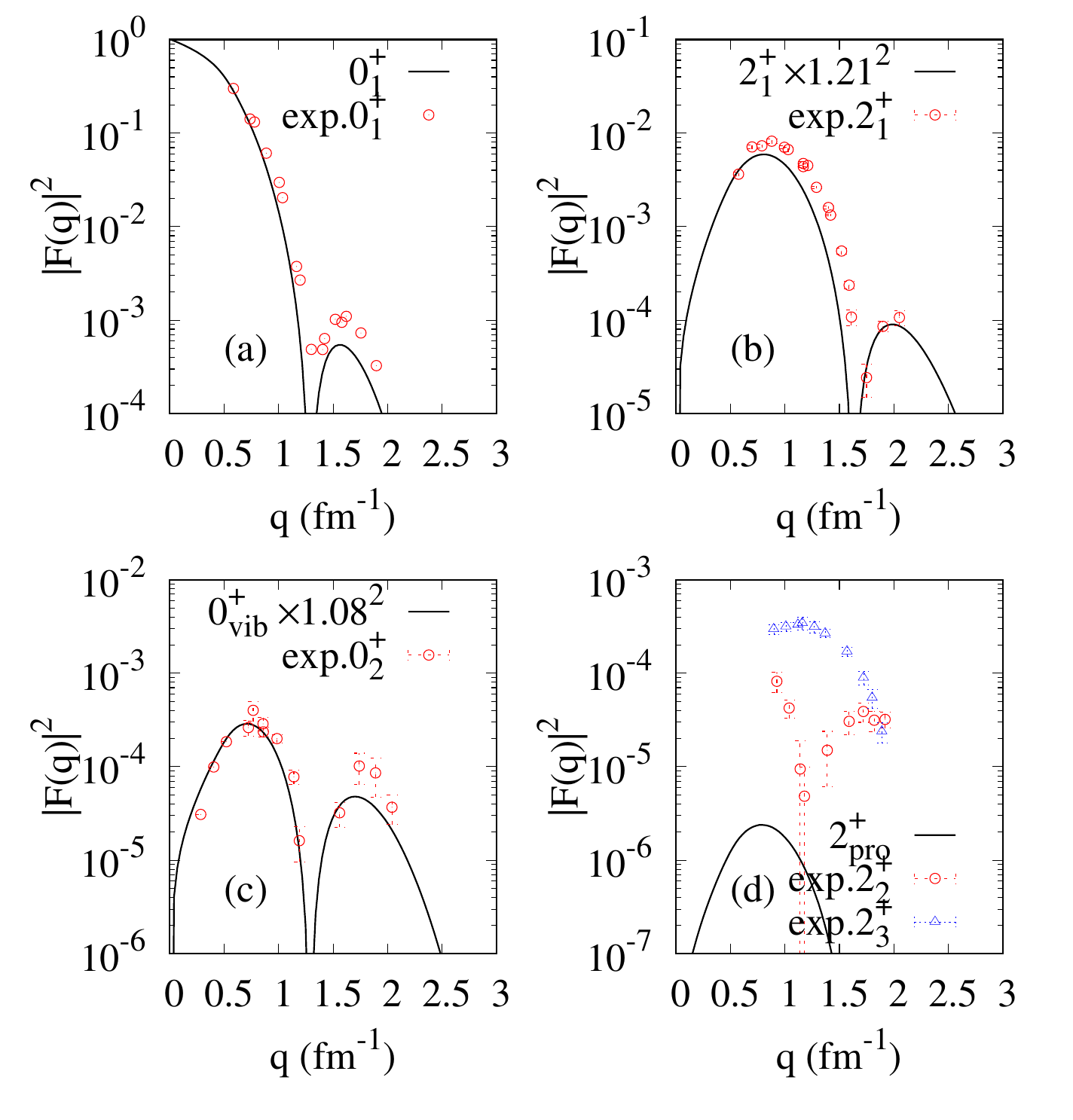}
  \caption{Elastic and inelastic form factors of  positive-parity states of $^{28}$Si.
The inelastic form factors obtained by AMD are renormalized 
by $f^2_\textrm{tr}$ with 
the factor ($f_\textrm{tr}$) listed in Table~\ref{tab:BEl}.
The experimental data are those measured by electron scattering 
in Refs.~\cite{Nakada:1972,Yen:1983sk,Chen:1990zza}.
  \label{fig:form-fig1}}
\end{figure}

\begin{figure}[!h]
\includegraphics[width=8.5 cm]{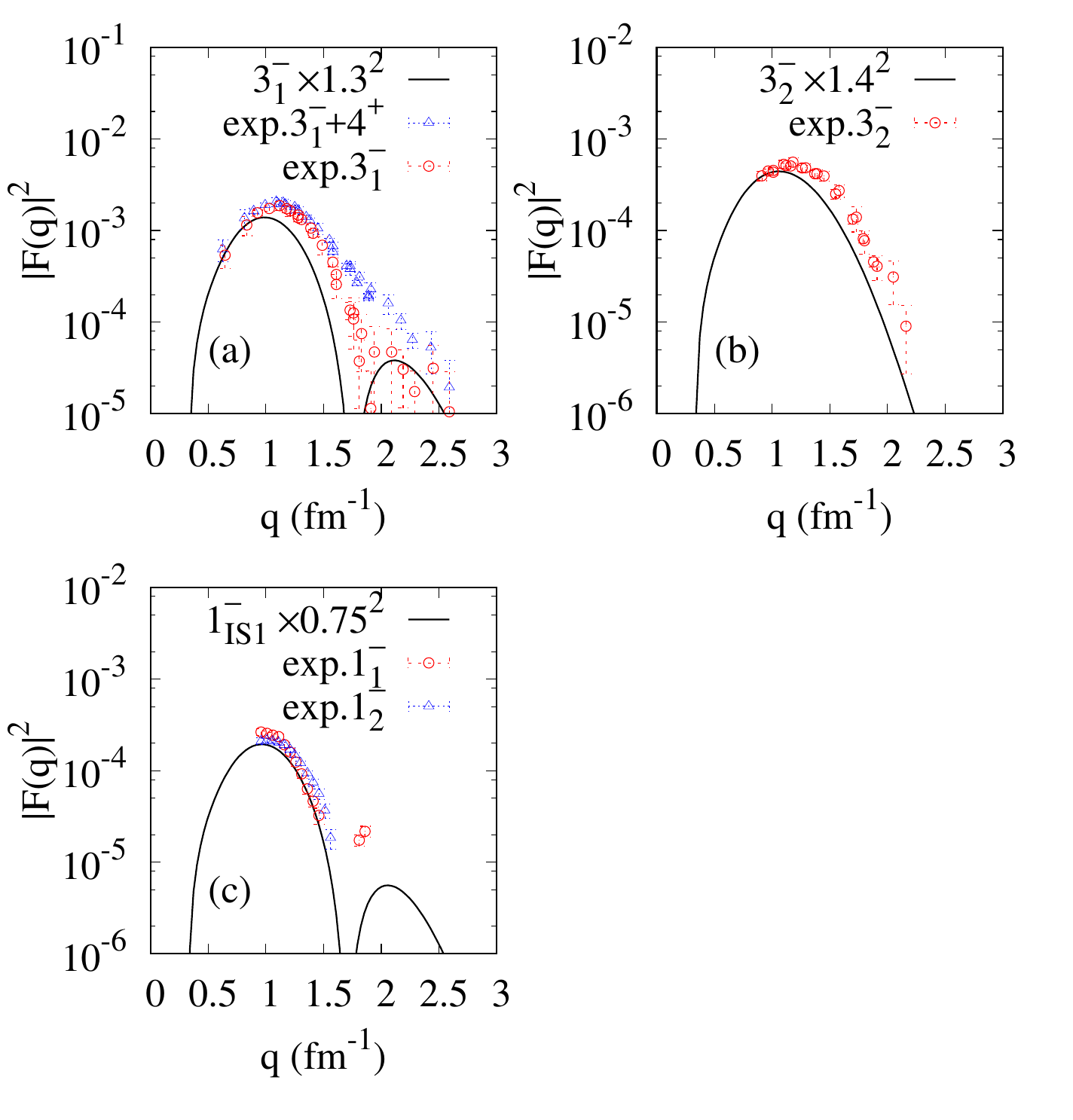}
  \caption{Inelastic form factors of negative-parity states of $^{28}\textrm{Si}$.
The inelastic form factors obtained by AMD are renormalized 
by $f^2_\textrm{tr}$ with 
the factor ($f_\textrm{tr}$) listed in Table~\ref{tab:BEl}.
The experimental data measured by electron scattering are from
Refs.~\cite{Nakada:1972,Yen:1983sk,Chen:1990zza}.
In the panel (a) for the  $3^-_1$ state, triangles indicate the experimental data of sum of the 
$3^-_1$(6.879 MeV) and $4^+_2$(6.888 MeV) contributions, and circles indicate the $3^-_1$(6.879 MeV) data
evaluated by subtracting the $4^+_2$(6.888 MeV) contribution from the sum \cite{Yen:1983sk}.
  \label{fig:form-fig2}}
\end{figure}

The renormalized form factors for positive- and negative-parity states 
are compared with experimental data in Figs.~\ref{fig:form-fig1} and \ref{fig:form-fig2}, respectively. 
The inelastic form factors of the $0^+_\textrm{vib}$, $2^+_1$, $3^-_1$, and $3^-_2$ states 
are reproduced reasonably by the calculation after the renormalization. 
For the $1^-$ state, the observed form factors of $1^-_1$ (8.95 MeV) and $1^-_2$ (9.93 MeV) measured by  $(e,e')$ expeirment
are similar to each other.   
The result of the  $1^-_\textrm{IS1}$ state is in reasonable agreement with 
the form factors of the two $1^-$ states, and suggests a possible assignment to either of these two 
states. 
For the $2^+_\textrm{pro}$ state, the calculation predicts considerable suppression of the 
inelastic transitions because of structure difference between the oblate and 
prolate bands.
However, 
the observed form factors of the $2^+_2$(7.32 MeV) state is larger by two orders of magnitude 
than the calculation. Not only the magnitude but also the $q$ dependence of the observed form factors 
are different from the calculation. It suggests that the prolate $2^+$ state may contain
significant mixing of other $2^+$ component beyond the present framework. 
In other words, the inelastic transition of $0^+_1\to 2^+_2$ 
probes the mixing component rather than the prolate $2^+$ component. It is contrast 
to the $E2$ transition from the $4^+_3$ state, which is dominantly contributed by 
the prolate $2^+$ component.

\begin{figure}[!h]
\includegraphics[width=7 cm]{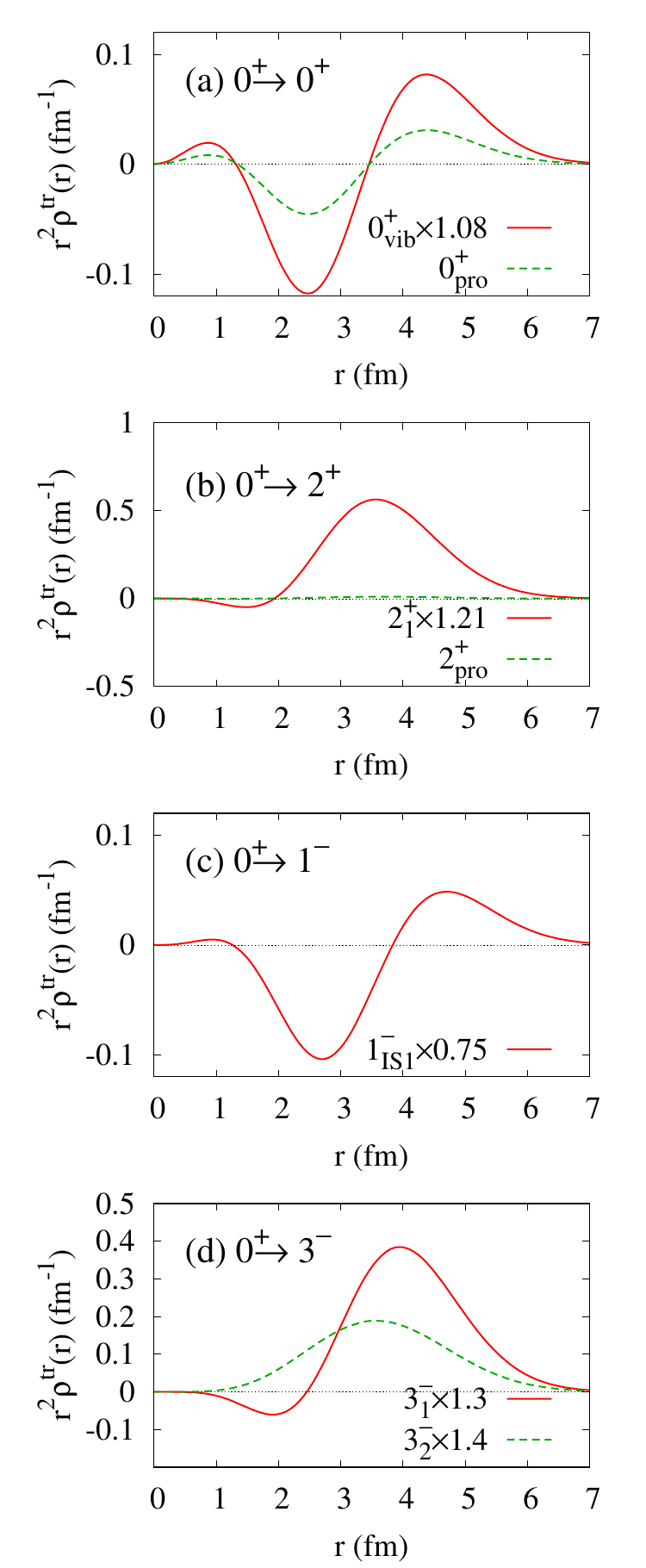}
  \caption{Transition densities of rank $\lambda=J$ transitions 
from the ground state to $J^\pi$ states. 
The theoretical values calculated with AMD are renormalized by $f_\textrm{tr}$ listed 
in Table~\ref{tab:BEl}.
\label{fig:trans}}
\end{figure}

The renormalized transition densities are shown in Fig.~\ref{fig:trans}. 
The transition densities from the ground state to the  $0^+_\textrm{vib}$ and $0^+_\textrm{pro}$ states (Fig.~\ref{fig:trans}(a))
show a similar $r$ behavior having two nodes at 
almost the same positions though the amplitude for the $0^+_\textrm{pro}$ is much smaller. 
In comparison of the transition densities for the $3^-_1$ and $3^-_2$ states,
one can see a quite different $r$ behavior between the two $3^-$ states (Fig.~\ref{fig:trans}(d)):
The transition density to the $3^-_1$ state
has a node at 2.5 fm and remarkable amplitudes in the outer region, while 
that to the $3^-_2$ has amplitudes in the inner region without node. 
This difference can be observed in the form factors 
shown in Fig. \ref{fig:form-fig2}. The calculated form factors show different dip positions 
between the $3^-_1$ and $3^-_2$ states, and seems consistent with the $(e,e')$ data.

\begin{table}[ht]
\caption{Excitation energies and root mean square radii of $^{28}$Si.
Calculated values obtained by AMD and experimental values are listed.
The theoretical $0^+_\textrm{vib}$ and $0^+_\textrm{pro}$ are assigned to the experimental 
$0^+_2$ and $0^+_3$ states. Assignment of the theoretical $2^+_\textrm{pro}$,  
$4^+_\textrm{pro}$,  $1^-_\textrm{IS1}$, and $3^-_2$ states are tentative.
The experimental energies are from Ref.~\cite{Basunia:2013}. 
The experimental data of the point-proton rms radius of the ground state is $R=3.010(24)$~fm 
from the experimental charge radius \cite{Angeli2013}.
 \label{tab:radii}
}
\begin{center}
\begin{tabular}{llllllrrrrccccc}
\hline
\hline
\multicolumn{2}{c}{exp}     &    \multicolumn{3}{c}{AMD} \\
  $J^\pi$  & $E_x$ (MeV) &  $J^\pi$     & $E_x$ (MeV) & $R$ (fm) \\
$0^+_1$	&	0	&	$0^+_1$	&	0.0 	&	3.17 	\\
$0^+_2$	&	4.98	&	$0^+_\textrm{vib}$	&	5.4 	&	3.17 	\\
$0^+_3$	&	6.691	&	$0^+_\textrm{pro}$	&	5.2 	&	3.31 	\\
$2^+_1$	&	1.779	&	$2^+_1$	&	2.1 	&	3.22 	\\
$2^+_2$	&	7.32	&	$2^+_\textrm{pro}$	&	6.0 	&	3.34 	\\
$2^+_3$	&	7.42	&		&		&		\\
$4^+_1$	&	4.618	&	$4^+_1$	&	4.1 	&	3.23 	\\
$(4^+_3)$	&	9.16	&	$4^+_\textrm{pro}$	&	7.8 	&	3.34 	\\
$1^-_1$	&	8.95	&	$1^-_\textrm{IS1}$	&	15.3 	&	3.29 	\\
$1^-_2$	&	9.93	&		&		&		\\
$3^-_1$	&	6.879	&	$3^-_1$	&	11.1 	&	3.28 	\\
$(3^-_2)$	&	10.18	&	$3^-_2$	&	16.7 	&	3.29 	\\
\hline
\hline
\end{tabular}
\end{center}
\end{table}

\begin{table}[ht]
\caption{
The $E\lambda$ ad IS1 transition strengths of $^{28}$Si
calculated with AMD and the experimental values 
measured by $\gamma$-decay life times 
and electron scattering.
For the IS1 transition strengths
of the $1^-\to 0^+$ transitions, the values of $B(\textrm{IS}1)/4$ are shown.
Values of the renormalization factor $f_\textrm{tr}$ are 
determined by the ratio of the experimental value $B_\textrm{exp}(E\lambda)$
to the calculated value $B_\textrm{cal}(E\lambda)$ as $f_\textrm{tr}=\sqrt{B_\textrm{exp}(E\lambda)/B_\textrm{cal}(E\lambda)}$.
For the $3^-_2\to 0^+_1$ transition, $f_\textrm{tr}=1.40$ is chosen
to fit the inelastic form factors of the $(e,e')$ data~\cite{Yen:1983sk}. 
The experimental data $B(E\lambda)$ are values reduced from $\gamma$-decay life times 
\cite{Basunia:2013} and the $(e,e')$ data \cite{Yen:1983sk,Chen:1990zza}. 
 \label{tab:BEl}
}
\begin{center}
\begin{tabular}{llllllcccc}
\hline
\hline
& \multicolumn{2}{c}{exp} &\multicolumn{3}{c}{AMD}    \\
    &     $B(E\lambda)$ \cite{Basunia:2013} &  $(e,e')$  \cite{Yen:1983sk,Chen:1990zza} &   & $B(E\lambda)$ & $f_\textrm{tr}$ \\
\multicolumn{2}{l}{$B(E0)$ ($e^2\textrm{fm}^{4})$} \\																	
$0^+_2\to 0^+_1$	&$					$&$	4.71 				$&	$0^+_\text{vib}\to 0^+_1$	&	4.0 	&	1.08 	\\
$0^+_3\to 0^+_1$	&$					$&$					$&	$0^+_\text{pro}\to 0^+_1$	&	0.7 	&		\\
\multicolumn{2}{l}{$B(\textrm{IS1})/4$ ($e^2\textrm{fm}^{6})$} \\
$1^-_1\to 0^+_1$	&$					$&$	18.7 				$&	$1^-_1\to 0^+_1$	&	34	&	0.75 	\\
\multicolumn{2}{l}{$B(E2)$ ($e^2\textrm{fm}^{4})$} \\
$2^+_1\to 0^+_1$	&$	67 	(	3 	)	$&$	55.7 				$&	$2^+_1\to 0^+_1$	&	46 	&	1.21 	\\
$2^+_2\to 0^+_1$	&$	1.87 	(	0.76 	)	$&$	1.26 				$&	$2^+_2\to 0^+_1$	&	0.03 	&		\\
$2^+_3\to 0^+_1$	&$	0.82 	(	0.09 	)	$&$	0.90 				$&		&		&		\\
$0^+_2\to 2^+_1$	&$	48 	(	3 	)	$&$					$&	$0^+_\text{vib}\to 2^+_1$	&	79 	&	0.78 	\\
$0^+_3\to 2^+_1$	&$	1.3 	(	0.1 	)	$&$					$&	$0^+_\text{pro}\to 2^+_1$	&	15.6 	&	0.29 	\\
$4^+_1\to 2^+_1$	&$	82.8 	(	9.1 	)	$&$					$&	$4^+_1\to 2^+_1$	&	87 	&		\\
$4^+_3\to 2^+_1$	&$	0.4 	(	0.1 	)	$&$					$&	$4^+_\textrm{pro}\to 2^+_1$	&	0.01 	&		\\
$4^+_3\to 2^+_2$	&$	152 	(	20 	)	$&$					$&	$4^+_\textrm{pro}\to 2^+_\textrm{pro}$	&	236 	&		\\
$4^+_3\to 2^+_3$	&$	56.1 	(	9.1 	)	$&$					$&		&		&		\\
$3^-_1\to 1^-_1$	&$					$&$					$&	$3^-_1\to 1^-_1$	&	4.7 	&		\\
$3^-_2\to 1^-_1$	&$					$&$					$&	$3^-_2\to 1^-_1$	&	75 	&		\\
\multicolumn{2}{l}{$B(E3)$ ($e^2\textrm{fm}^{6})$} \\													
$3^-_1\to 0^+_1$	&$	615 	(	70 	)	$&$	553 	(	107 	)	$&	$3^-_1\to 0^+_1$	&	366 	&	1.30 	\\
$3^-_2\to 0^+_1$	&$					$&$	78 	(	20 	)	$&	$3^-_2\to 0^+_1$	&	76&	1.40\\
\multicolumn{2}{l}{$B(E0)$ ($e^2\textrm{fm}^{4})$} \\																	
$4^+_1\to 0^+_1$	&$					$&$	2734 				$&	$4^+_1\to 0^+_1$	&	2500 	&		\\
\hline
\hline
\end{tabular}
\end{center}
\end{table}

\section{Results of proton and $\alpha$ scattering}  \label{sec:results2}

The MCC calculations of proton scattering at incident energies $E_p=65$, $100$, and $180$ MeV and 
$\alpha$ scattering at incident energies $E_\alpha=120$, 130, 240, and 400 MeV
are performed using the matter and renormalized transition densities obtained by AMD. 
In the MCC calculations, we take into account 
$\lambda=0,1,2,3$ transitions between  
the $0^+_1$, $0^+_\textrm{vib}$, $0^+_\textrm{pro}$, $1^-_\textrm{IS1}$, $2^+_1$,  $2^+_\textrm{pro}$, $3^-_1$, and $3^-_2$ states, and use the experimental excitation energies following the assignments to 
the $0^+_1$, $0^+_2$(4.98 MeV), $0^+_2$(4.98 MeV), $2^+_1$(1.779 MeV), $2^+_2$ (7.32 MeV), 
$1^-_1$ (8.95 MeV), $3^-_1$ (6.879 MeV), and $3^-_2$ (10.18 MeV) states, respectively.
To see the CC effect, the one-step calculation of the distorted wave born approximation (DWBA) is 
also performed. 
In the following discussions of the calculated  cross sections,
we use labels of $0^+_{1,2,3}$, $1^-_1$, $2^+_{1,2}$, and $3^-_{1,2}$ corresponding to the 
above assignments unless otherwise noted. 

\subsection{Elastic scattering}

\begin{figure}[!h]
\includegraphics[width=8 cm]{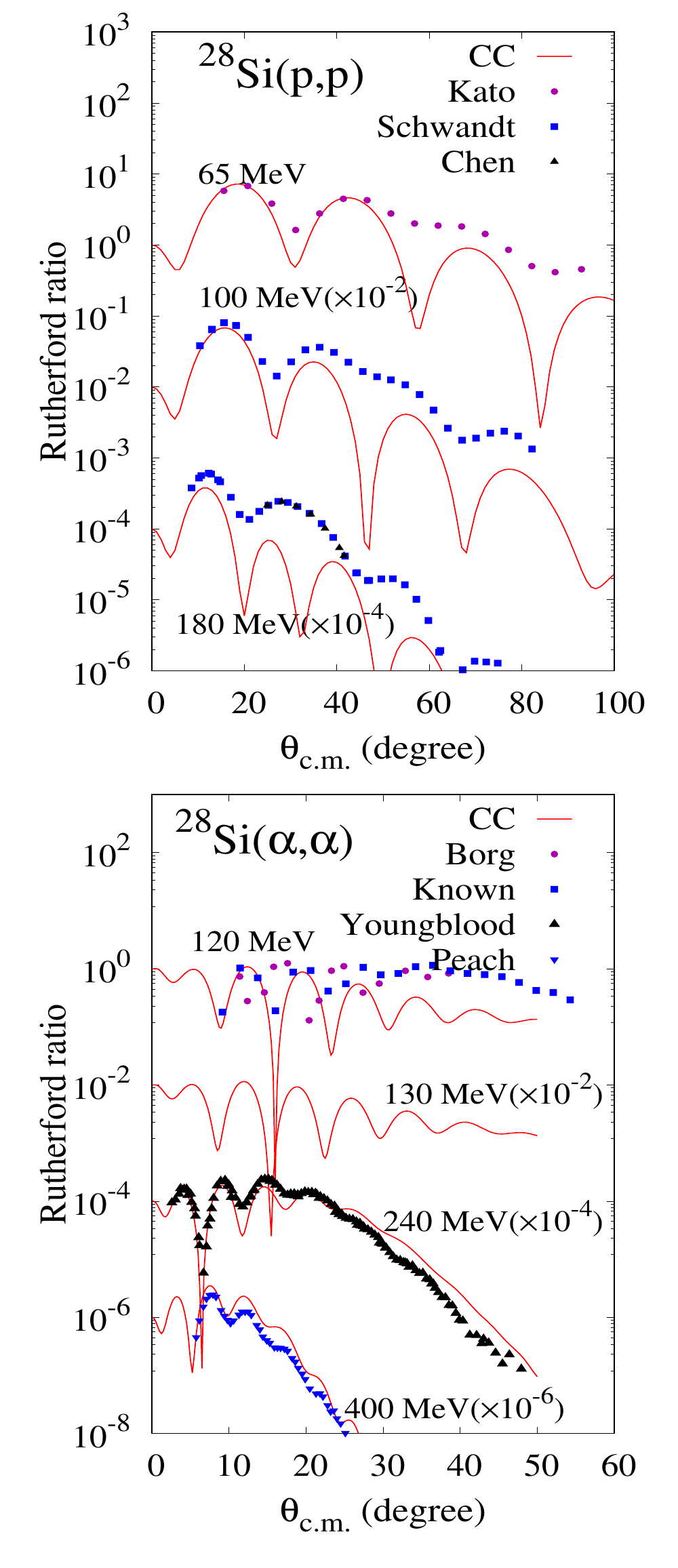}
\caption{Cross sections of elastic proton and $\alpha$ scattering off $^{28}$Si calculated
with the CC calculation for proton
incident energies $E_p=65$, 100, 180 MeV  and $\alpha$ incident energies 
$E_\alpha$=120, 130, 240, and 386 MeV.
The experimental data are $(p,p)$ cross sections at $E_p=65$ MeV\cite{Kato:1985zz}, $100$ MeV\cite{Schwandt:1982py}, and 180 MeV\cite{Chen:1990zza}, and 
 $(\alpha,\alpha)$ cross sections at $E_\alpha=120$ MeV\cite{VanDerBorg:1981qiu,Kwon:2007}, 
240 MeV\cite{Youngblood:2002mk}, and 386 MeV\cite{Peach:2016yop}. 
  \label{fig:cross-pa}}
\end{figure}

In Fig.~\ref{fig:cross-pa},
the elastic proton and $\alpha$ scattering cross sections are shown compared with the experimental data.
The calculation reasonably reproduces amplitudes of the $(p,p)$ cross sections at $E_p=65$, 100, and 180 MeV
and qualitatively describes diffraction patterns,
though it is not precise enough to reproduce dip structure at large angles mainly because the 
spin-orbit potentials are ignored in the present calculation.
At higher energies, even the cross sections around the peaks are undershot for the same reason.
For $\alpha$ scattering, the calculation successfully reproduces
amplitude and diffraction patterns of the elastic cross sections at $E_p=240$ and 400 MeV.
For lower energies, agreement with the data is reasonable but 
the observed data are not enough precise for detailed discussions and even inconsistent between different experiments.

\subsection{Inelastic scattering of proton}
\begin{figure*}[!h]
\includegraphics[width=17 cm]{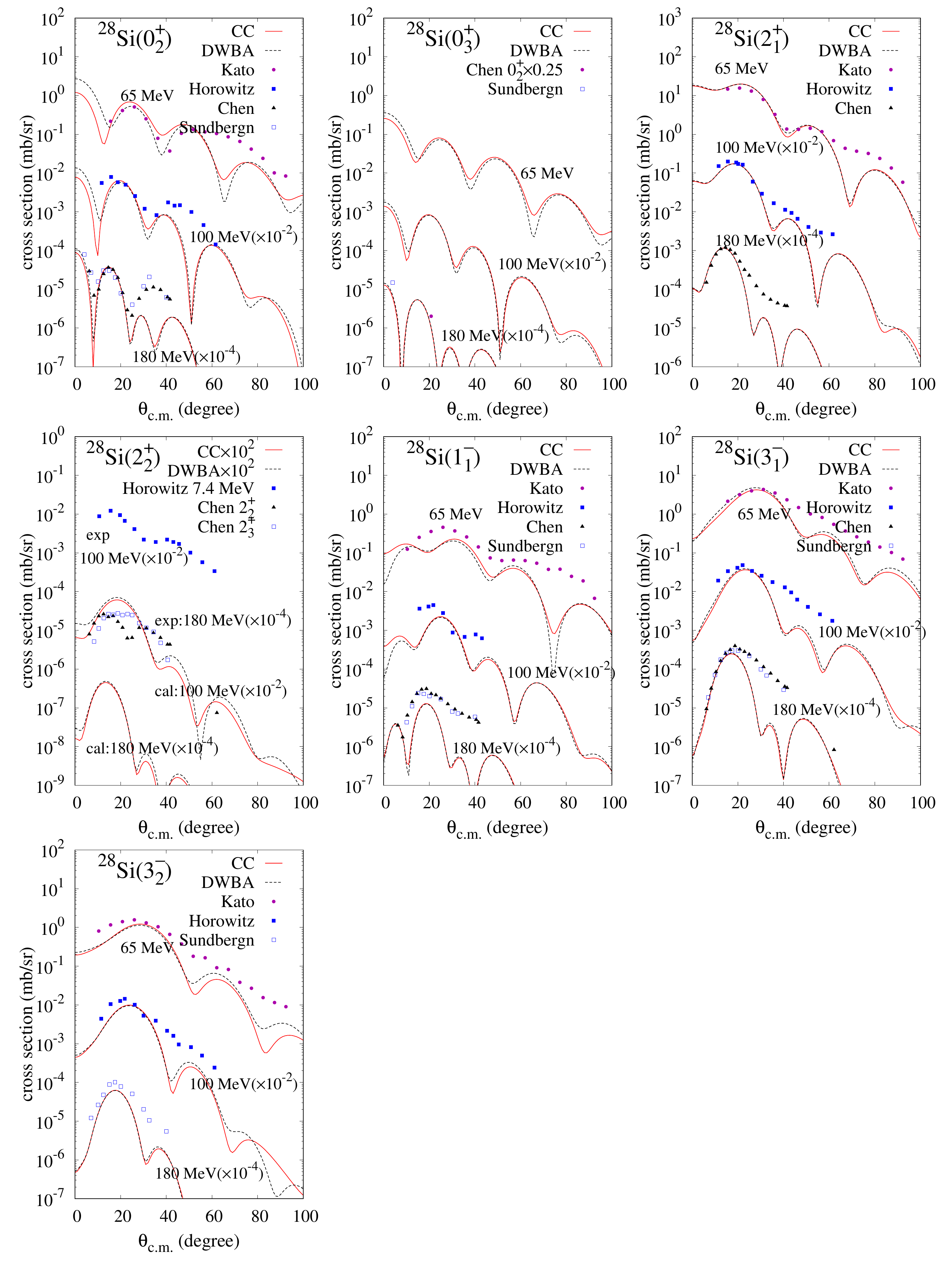}
\caption{Proton inelastic scattering cross sections at  incident energies $E_p=65$, 100, 180 MeV obtained by 
the CC and DWBA calculations. 
Experiment data are cross sections at $E_p=65$~MeV\cite{Kato:1985zz},  $E_p=100$~MeV\cite{Horowitz:1969eso}, and 
 $E_p=180$~MeV\cite{Chen:1990zza},  and $E_p=185$~MeV\cite{Sundberg:1967rjo}.
For the experimental data of the $0^+_3$ cross sections, 
the upper limit of the  $0^+_3$ cross section at 
$\theta_\textrm{c.m.}=4^\circ$ from Ref.~\cite{Sundberg:1967rjo} and 
a quoter of the $0^+_2$ cross section. at $\theta_\textrm{c.m.}=20^\circ$
evaluated from the observed spectrum shown in Ref.~\cite{Chen:1990zza}
are shown. See text. 
 \label{fig:cross-si28p}}
\end{figure*}

Figure \ref{fig:cross-si28p} shows the $(p,p')$ cross sections of the
$0^+_{2,3}$, $2^+_{1,2}$, $1^-_1$, and $3^-_{1,2}$ states. 
Results of the CC (solid lines) and DWBA (dashed lines) calculations
are shown 
together with the experimental data. One can see that the CC effect in proton scattering is generally minor 
in this energy range.
For the $0^+_2$, $2^+_1$, $3^-_1$, 
and $3^-_2$ states, 
the CC calculation reasonably reproduces 
amplitudes and diffraction patterns of the $(p,p')$ cross sections 
at forward peaks.
It also describes the observed $1^-_1$ cross sections qualitatively, but the agreement with the 
experimental data is not satisfactory. For the $2^+_2$ state, 
the calculation fails to reproduce the experimental data: the calculated cross sections are
smaller than the data by two orders of magnitude, consistently with the underestimation of the form factors. 
This result suggests again possible mixing of other component with the prolate component
in the $2^+_2$ state.
As for the $0^+_3$ state, there is only a few data of $(p,p')$ cross sections at $E_p=180$--$185$ MeV.
In the $(p,p')$ experiment at $E_p=180$ MeV, 
weak production of the
$0^+_3$ state has been observed. From the peak hight in the observed spectrum shown in 
Fig.~1 of Ref.~\cite{Chen:1990zza}, one can roughly estimate the $0^+_3$ cross section 
at $\theta_\textrm{c.m.}=20^\circ$ as 1/4 of the $0^+_2$ cross section. 
In the experiment at $E_p=185$ MeV\cite{Sundberg:1967rjo}, 
the upper limit of the  $0^+_3$ cross section at 
$\theta_\textrm{c.m.}=4^\circ$ was reported.
These two data are plotted for order estimation of the $0^+_3$ cross sections 
in Fig.~\ref{fig:cross-si28p}. The data seem to be consistent with the 
calculated $0^+_3$ cross sections at $E_p=180$ MeV, but quality of the data is not enough to 
clarify the transition properties of the $0^+_3$ state. 

\subsection{Inelastic scattering of $\alpha$}
\begin{figure*}[!h]
\includegraphics[width=17 cm]{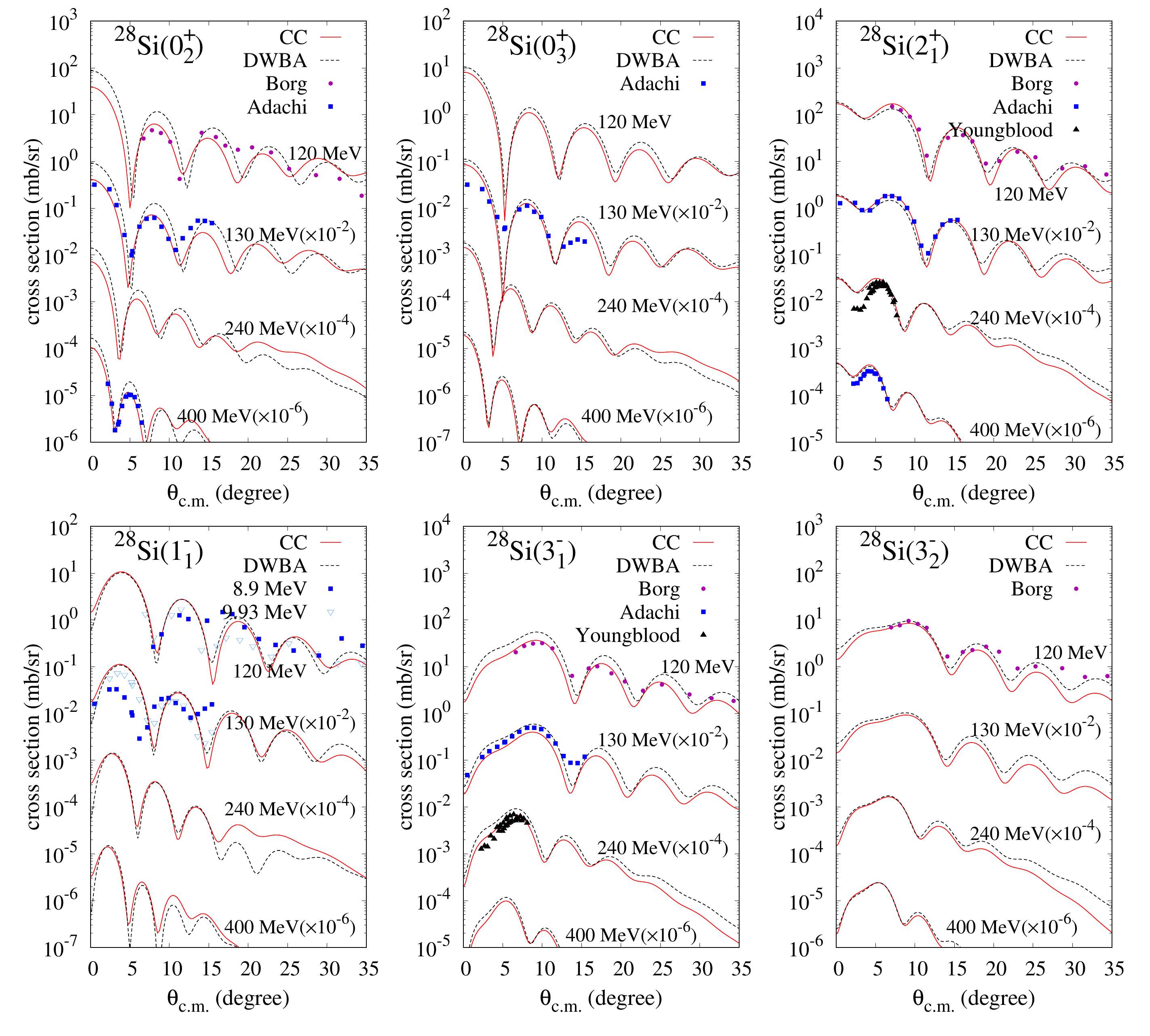}
\caption{$\alpha$ inelastic scattering cross sections at  incident energies $E_\alpha=120$, 130, 240, and 400 MeV obtained by 
the CC and DWBA calculations. 
Experiment data are $(\alpha,\alpha')$ cross sections at $E_\alpha=120$ MeV~\cite{VanDerBorg:1981qiu}, 
130 MeV~\cite{Adachi:2018pql}, 
240 MeV~\cite{Youngblood:2002mk},  and 386~MeV\cite{Adachi:2018pql}.
  \label{fig:cross-si28a}}
\end{figure*}

The calculated cross sections for $\alpha$ inelastic scattering at $E_\alpha=$120, 130, 240, and 400
MeV are shown in Fig.~\ref{fig:cross-si28a} compared with experimental data for 
$E_\alpha=120$~\cite{VanDerBorg:1981qiu}, 
 130~\cite{Adachi:2018pql}, 
 240~\cite{Youngblood:2002mk},  and 386~MeV~\cite{Adachi:2018pql}.
One can see that 
the CC effect is significant in the  $0^+_2$ state and non negligible in the  $3^-_1$ state, while 
it is relatively minor in the  $0^+_3$, $2^+_1$, $1^-_1$, and  $3^-_2$ cross sections.
The CC effect becomes weaker as the incident energy increases as expected. 
The CC calculation successfully reproduces 
the  $0^+_2$, $2^+_1$, and $3^-_1$ cross sections with good description of 
amplitudes and diffraction patterns in a wide energy range.
It also describes well the experimental cross sections of the $0^+_3$ state at $E_p=130$ MeV
and those of the $3^-_2$ state at $E_\alpha=120$ MeV.
These results support validity of the present MCC approach and accuracy 
of the adopted transition densities. 
For $1^-$ states, the calculated $1^-_1$ cross sections reasonably agree with the experimental 
cross sections of the $1^-_1$ (8.9 MeV) state and also coincides with those of the 
$1^-_2$ (9.93 MeV) state. However, the experimental data are available only for low incident energies and 
not enough to draw a definite conclusion for assignment of the theoretical $1^-$ state. 

\section{Discussion} \label{sec:discussions}
In the previous sections, we showed the calculated results of form factors, $p$ scattering, and 
$\alpha$ scattering, and compared them with the observed data.
In this section, we discuss transition properties of excited states by combining these results of  
inelastic scattering as well as structure features such as transition strengths.

For the $0^+_2$, $2^+_1$, and $3^-_1$ states, details of transition properties such as 
$E\lambda$ transition strengths and form factors are experimentally known.
The present calculation reasonably reproduces the experimental values of $B(E\lambda)$.
After fine tuning by the renormalization, the experimental form factors are described well by the 
calculation. The MCC calculation with renormalized transition densities 
reproduces successfully $\alpha$ inelastic scattering in a wide energy range 
and reasonably describes the observed data of proton inelastic scattering.
It should be stressed that we can obtain consistent results for electric and hadron scattering
within a microscopic framework and confirm 
the applicability of the present MCC approach.
Combining the structure analysis, these states are understood as  
the $0^+$, $2^+$, and $3^-$ excitations built on the 
oblate ground state. 

For the $0^+_3$ state, there is no experimental information from electric probes 
such as form factors. 
Using the calculated transition density, which gives the strength 
$B(E0; 0^+_1\to 0^+_3)=0.7$ $e^2$fm$^4$, 
the present MCC calculation
reproduces the $\alpha$ inelastic cross sections of the $0^+_3$ state. 
It gives a result being consistent with the experimental observations of proton inelastic scattering.
It should be commented again that the predicted transition strength 
may contain model ambiguity from state mixing of the vibration and prolate $0^+$ modes.
However, radial behavior ($r$ dependence) of the $0^+$ transition density is rather 
stable against such the state mixing, and therefore, the ambiguity 
may exist only in the overall factor of the $0^+_1\to 0^+_3$ transition density. 
From the successful result for reproduction of the $\alpha$ inelastic scattering data, 
we can say that the predicted value $B(E0; 0^+_1\to 0^+_3)=0.7$ $e^2$fm$^4$ is likely to be reasonable.

For $1^-$ states, inelastic form factors have been experimentally observed for the $1^-_1$ (8.9 MeV) and 
$1^-_2$ (9.93 MeV) states, but low $q$ data are not enough to 
determine the IS1 transition strength with high precision. 
In the structure calculation of AMD, 
the $1^-_\textrm{IS1}$ is obtained and regarded as the 
IS1 mode induced by $\alpha$-cluster excitation on the oblate ground state. 
Since the calculated form factors are consistent with the experimental data observed for the 
$1^-_1$ (8.9 MeV) and $1^-_2$ (9.93 MeV) states, 
a possible assignment of this IS1 mode is 
either of the experimental $1^-$ states.
Another possibility is that 
the IS1 mode is fragmented into the two $1^-$ states. 
The MCC calculation qualitatively describes proton scattering data of the $1^-_1$ (8.9 MeV) state
but the agreement is not enough accurate.
As for $\alpha$ scattering, 
the calculated cross sections of the  $1^-_\textrm{IS1}$ state 
are in reasonable agreement with the $(\alpha,\alpha')$ 
data of the $1^-_1$ (8.9 MeV) state and also with the data of the 
$1^-_2$ (9.93 MeV) state. In the present analysis, we can not conclude which assignment is more likely.

The $3^-_2$ state is obtained as a member of the $K^\pi=0^-$ band built on 
the $1^-_\textrm{IS1}$ state in the present calculation.
The calculated form factors can be adjusted to the observed data 
of the $3^-_2$ (10.18 MeV) with a renormalization factor.
For proton and $\alpha$ inelastic scattering, 
the $3^-_2$ cross sections obtained by the MCC calculation
correspond well to the experimental
cross sections observed for the  $3^-_2$ (10.18 MeV) state.
From this correspondence, the $3^-_2$ (10.18 MeV) is 
considered to be a member of the $K^\pi=0^-$ band 
constructed on the IS1 mode, which generated by the $\alpha$-cluster excitation 
on the ground state. 

The present calculation describes the shape coexistence of the 
oblate and prolate deformations.
The prolate deformation constructs the rotational band of the $0^+_\textrm{pro}$, 
$2^+_\textrm{pro}$, and $4^+_\textrm{pro}$ states. 
In the experimental energy spectra, there are two 
candidates for the prolate $2^+$ state as the $2^+_2$ (7.32 MeV) and $2^+_3$ (7.42 MeV)  states. 
The calculated  $B(E2;4^+_\textrm{pro}\to 2^+_\textrm{pro})$ of the in-band transition
agrees well with a sum of the observed strengths 
$B(E2;4^+_3\to 2^+_2)$ and  $B(E2;4^+_3\to 2^+_3)$, suggesting that 
the prolate $2^+$ state is likely to be fragmented into two $2^+$ state
via mixing with other $2^+$ component. 
In the present calculation for the $2^+_\textrm{pro}$ state, inelastic transitions 
from the ground state are strongly suppressed because of 
the shape difference between the initial and final states. 
As a result, weak form factors and inelastic scattering are obtained 
for the $2^+_\textrm{pro}$ state.  However, the observed 
form factors and proton scattering cross sections of the $2^+_2$ state 
are considerably large as two orders of magnitudes as the calculation.
In other words, the significant form factors and proton inelastic cross sections can be understood as 
experimental signals of mixing of the other $2^+$ component, which is beyond the present 
structure model calculation. 

\section{Summary} \label{sec:summary}
Transition properties of $0^+$, $1^-$, $2^+$, and $3^-$ states of $^{28}\textrm{Si}$ 
were investigated via proton and $\alpha$ inelastic scattering. 
The structure calculation was performed with 
the energy variation after total angular momentum and parity projections in the AMD 
framework. In the AMD calculation, the oblate ground and prolate bands,
$0^+$ and $3^-$ excitations, and the $1^-$ and $3^-$ states of the $K^\pi=0^-$ band were
obtained. The calculation reasonably reproduced the transition properties 
such as transition strengths and form factors.

Using the matter and transition densities of  $^{28}\textrm{Si}$ obtained by the AMD calculation, 
the MCC calculations of proton and $\alpha$ scattering off $^{28}\textrm{Si}$ are
performed. The proton-$^{28}\textrm{Si}$ and $\alpha$-$^{28}\textrm{Si}$ potentials are microscopically
derived by folding the Melbourne $g$-matrix $NN$ interaction with the 
 $^{28}\textrm{Si}$ and $\alpha$ densities. In order to reduce possible ambiguity from the structure model, 
the theoretical transition densities were renormalized to 
fit the $B(E\lambda)$ for the use of the MCC calculation.
The MCC calculation reasonably reproduces the observed elastic and inelastic cross sections of 
proton and $\alpha$ scattering. From the analysis of inelastic scattering 
combined with structure properties, we assigned the theoretical states to the observed levels. 

The $2^+_1$, $0^+_2$, and $3^-_1$ states 
are understood respectively as the oblate ground band member, vibration $0^+$ and $3^-$ excitations 
built on the oblate ground state.
The MCC calculation reproduces well the proton and $\alpha$ inelastic cross sections of these states 
in a wide energy range of $E_p=65$--180 MeV
and $E_\alpha=120$--400 MeV.
It should be stressed that 
consistent results for electron, proton and $\alpha$ scatterings are obtained 
within a microscopic framework. These results proved applicability of the present MCC approach
for proton and $\alpha$ inelastic processes.

For the $0^+_3$ state in the prolate band, 
the calculated $E0$ transition strength is relatively weak 
compared with that of the $0^+_2$ state because of the shape difference between 
the oblate ground (initial) and prolate (final) states.
The predicted strength, $B(E0; 0^+_1\to 0^+_3)=0.7$ $e^2$fm$^4$, is 
supported by the observed cross sections of $\alpha$ scattering. 
It is also consistent with the proton scattering.

In the structure calculation, the $K^\pi=0^-$ band of the $1^-$ and $3^-$ states 
is constructed from the IS1 mode, which 
is induced by the $\alpha$-cluster excitation on the oblate ground state. 
From the analysis of form factors and inelastic scattering, 
the $3^-$ state corresponds to
the $3^-_2$ (10.18 MeV) in the experimental spectra.
The $1^-$ state is likely to be assigned to either of 
the $1^-_1$ (8.9 MeV) and $1^-_2$ (9.93 MeV) states, but we can not draw a conclusion
in the present analysis.

An advantage of the present MCC approach is that we can discuss electron, 
proton, and  $\alpha$ inelastic scattering within a unified treatment of microscopic descriptions.
Another merit is that there is no adjustable parameter in the reaction part. 
For given densities of a target nucleus, we can obtain 
the $(p,p')$ and $(\alpha,\alpha')$ cross sections at given energies without parameter tuning.
Owing to such the straightforward connection between structure inputs and output cross sections, 
validity of a structure input can be examined via proton and $\alpha$ cross sections,
even if electric data are not accurate enough.
It has been proved that analysis of proton and $\alpha$ inelastic scattering 
with the MCC calculation using the microscopic structure calculation 
is a useful tool to investigate properties of excited states.

\begin{acknowledgments}
The computational calculations of this work were performed by using the
supercomputer in the Yukawa Institute for theoretical physics, Kyoto University. This work was partly supported
by Grants-in-Aid of the Japan Society for the Promotion of Science (Grant Nos. JP18K03617, JP16K05352, and 18H05407) and by the grant for the RCNP joint research project.
\end{acknowledgments}

\end{document}